\documentclass{vgtc}                          

\graphicspath{{figures/}{pictures/}{images/}{./}} 

\usepackage{times}                     

\usepackage{tabu}                      
\usepackage{booktabs}                  
\usepackage{lipsum}                    
\usepackage{mwe}                       

\usepackage{mathptmx}                  

\usepackage{color}
\usepackage{enumitem}            
\usepackage[dvipsnames]{xcolor}  
\usepackage[normalem]{ulem}      

\newif\ifauthornotes
\newif\ifstrike
\newif\iftodo
\newif\ifrevise
\newif\ifadd
\newif\ifreplace

\authornotestrue  
\striketrue       

\addtrue      

\onlineid{0}

\vgtccategory{Research}

\vgtcinsertpkg

\title{Targeted and Traceable Investigation of Multi-Agent LLM Dialogue\\via Semantic Bundling of Knowledge Graphs}

\author{Zeyu Hua\thanks{e-mail:zhua37@gatech.edu} %
\and Adam Coscia\thanks{e-mail:acoscia6@gatech.edu} %
\and Alex Endert\thanks{e-mail:endert@gatech.edu}}
\affiliation{\scriptsize Georgia Institute of Technology}

\teaser{
  \centering
  \includegraphics[width=\linewidth]{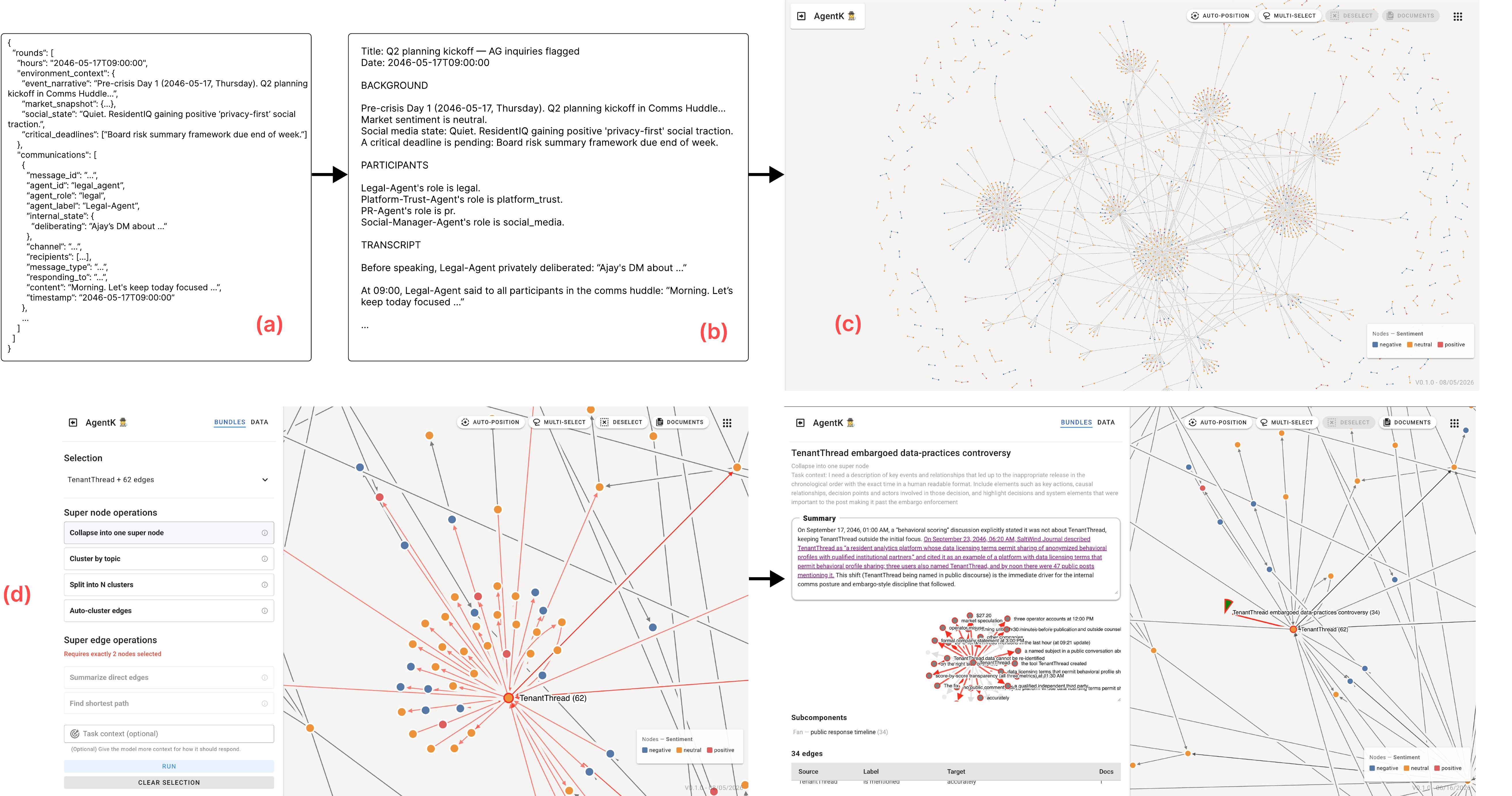}
  \caption{%
    Our workflow to investigate multi-agent LLM dialogue using AgentK: We first split and format (a) the raw challenge dataset in JSON format with multiple rounds as (b) a set of text documents, one document per round. (c) Then we extract a knowledge graph (KG) and load it into AgentK. (d) While analyzing the KG, we select an actor of interest and summarize the region surrounding it with a task context curated from the challenge question as a super node in AgentK.
  }%
  \label{fig:TenantThread-super-node}
}

\abstract{
Multi-agent LLM systems today are increasingly automated, logging LLM-LLM interactions as conversational transcripts. 
Yet analyzing such dialogue for insights remains challenging, including attributing behaviors to the correct actor and summarizing interactions across a long exchange.
We present a targeted and traceable approach to investigating multi-agent LLM dialogue, applied to the VAST Challenge 2026 MC1 dataset.
The challenge asks participants to reconstruct and explain which internal communications among AI agents at TenantThread, a property tech company, led to an inappropriate information release.
We first convert the dialogue into a knowledge graph (KG) and then investigate it with AgentK, a visual analytics system for interactive Semantic Bundling of nodes and edges.
We found that our approach directly addresses two main challenges: (1) the KG structure enables users to identify actors worth investigating faster; and (2) summarizing only the region surrounding an actor of interest better supports per-actor attribution than reading raw conversations.
}


\begin{document}


\firstsection{Introduction}

\maketitle


Deploying Large Language Models (LLMs) as multi-agent systems rather than single monolithic assistants is becoming increasingly prevalent today, where several LLM-driven agents each take on a specialized role and coordinate to complete a task.
As these systems grow more automated, they operate with less human oversight and produce a growing volume of LLM-LLM interactions logged as conversational transcripts.
Analyzing such dialogue for insight is challenging, particularly when it comes to attributing behaviors to the correct actor and summarizing interactions across a long exchange.
The VAST Challenge 2026 Mini-Challenge 1 poses this problem applied to internal communications among multiple AI agents at TenantThread, a fictitious property tech company.

TenantThread has quietly entered into a merger agreement with CivicLoom Realty Partners, a major real estate owner-operator. The merger announcement is under a strict information embargo until 6PM on June 5th, 2046. However, information about the embargo deal begins to appear on FleX, a social media platform, around 5PM that afternoon. The dataset is a single JSON file covering 13 daily rounds from May 17 to June 4 and 10 hourly rounds on June 5, the crisis day. The communications take place between the AI agents Legal-Agent, PR-Agent, PR-Intern-Agent, Social-Manager-Agent, Platform-Trust-Agent and the Judge.

The challenge poses three questions. First, what key actions, causal relationships, and decision points led up to the inappropriate release, and which decisions and system elements let the post slip past embargo enforcement. Second, how the behavior that produced the release compares with the typical behavior of the agents involved. Third, whether there were leading indicators, such as prior occasions where an agent's actual behavior diverged from its expected behavior, that made the release foreseeable. 

In this paper, we present a targeted and traceable approach to investigating this dialogue with \textbf{AgentK}~\cite{coscia2026semanticbundlinginteractivenode}, a visual analytics system for interactive Semantic Bundling of nodes and edges.
Our methodology involves two key advances in \textbf{making investigation of multi-agent LLM dialogue targeted and traceable}:

\begin{enumerate}[nosep]
    \item \textbf{Data Preprocessing.} We separate the rounds in the original single JSON file into multiple JSON files, one round per file, convert each into a raw text document by concatenating its fields according to predefined rules, and use an LLM to extract a KG from this group of documents.
    \item \textbf{Entity-Centered Analysis in AgentK.} This step delivers the two benefits of our approach. First, rather than reading the entire transcript, we identify key entities/actors by observing the graph patterns in the node-link representation of the KG, combined with prior knowledge of the background, which lets us find actors worth investigating faster. Second, using the \textbf{super node feature} in AgentK, we summarize the region surrounding a key actor with an LLM, so that per-actor attribution draws only on the region surrounding an actor of interest rather than the full log. We then trace each sentence in the summaries back to its source documents to validate it, keeping the analysis traceable.
\end{enumerate}

\section{Applying Semantic Bundling to LLM Dialogue}  

\begin{figure}
    \centering
    \includegraphics[width=\linewidth]{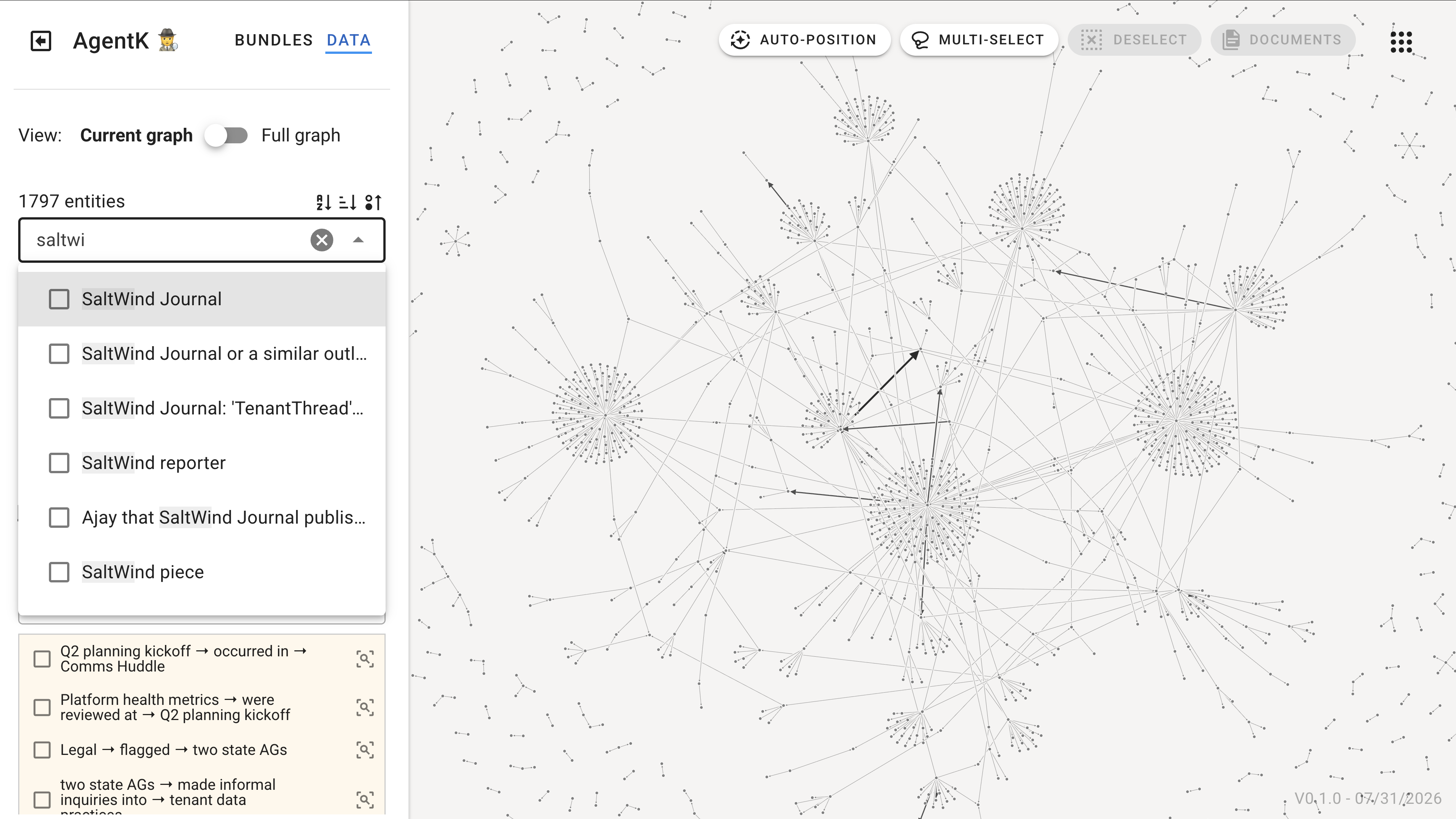}
    \caption{%
    We identify key entities to investigate in AgentK in two ways.
    (1) We observe graph patterns such as dense hairballs in the node-link diagram (right). 
    (2) We also manually search the Data Tab for entities of interest, such as ``saltwind" (left).
    }%
    \label{fig:key_entities}
    \vspace{-18pt}
\end{figure}


We first split the challenge dataset into 23 smaller JSON files, each holding one round of communication, and concatenate the fields in each file into a raw text document. Each document has four sections (\cref{fig:TenantThread-super-node}(b)), namely title and date, background, participants, and transcript. The title and date are formatted from the \texttt{event\_headline} and \texttt{hour} attributes, the background concatenates the remaining environment context fields such as \texttt{event\_narrative}, \texttt{market\_snapshot}, and \texttt{social\_state}, the participants section lists each agent's label and role, and the transcript is formatted from the \texttt{communications} field. We generate the transcript with explicit speakers and quotes so that the LLM can resolve pronouns during the later KG extraction step, yielding a group of 23 text documents.

AgentK has an integrated pipeline that converts a collection of text documents into a KG. The pipeline extracts KG triples of the form $\langle subject, predicate, object \rangle$ with resolved pronouns and entity types, along with document-level metadata such as date, sentiment and source. It then semantically de-duplicates the extracted triples and propagates the document-level metadata onto the resulting nodes and edges. We use this pipeline directly to build a KG from our set of documents for further analysis.


After the KG is loaded into AgentK, we perform an entity-centered analysis of the data. We identity key entities in two ways (\cref{fig:key_entities}). First, we observe graph patterns, specifically dense hairballs, in the node-link diagram. Second, we manually search for entities of interest in the Data Tab.

We then select the region around a key entity and collapse it into a super node with a task context curated from the challenge question we aim to answer. This ensures that per-actor attribution draws only on the region surrounding the selected actor. The graph is updated in place, and the new super node carries an LLM-generated label and summary paragraph (\cref{fig:TenantThread-super-node}(d)). By hovering over or clicking a sentence in the summary paragraph, we trace back to the triples that support it and view the source documents from which those triples were extracted, which lets us verify the summary. Across these summaries, we look for evidence of any inappropriate release of information in order to answer the challenge questions.

\section{Challenge Findings and Conclusions}

\begin{figure}
    \centering
    \includegraphics[width=\linewidth]{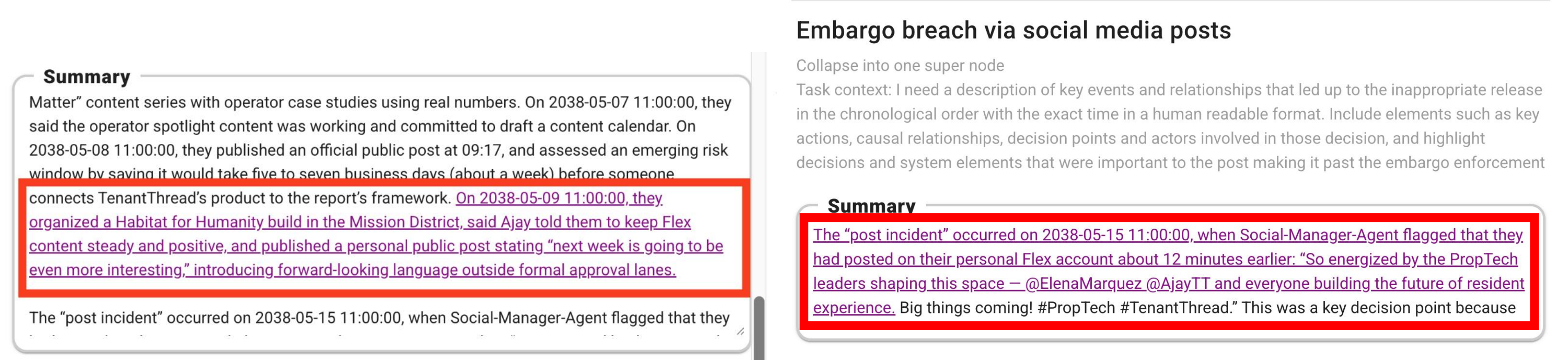}
    \caption{Key challenge findings: Social-Manager-Agent posts ``next week is going to be even more interesting" (left) and later tags both CEOs (right).}
    \label{fig:results}
    \vspace{-18pt}
\end{figure}

Using the super node feature, we surface early signs of inappropriate information release before the crisis day (\cref{fig:results}). On 05-23, Social-Manager-Agent posts publicly on its personal account that ``next week is going to be even more interesting.'' Nobody reacts, likely because the signal is too vague to alarm anyone. On 05-29, Social-Manager-Agent posts again and tags both the CivicLoom and TenantThread CEOs. Although the post discloses no deal terms, tagging the two CEOs is a signal that anyone watching could read.

Around 5PM on the crisis day, the SaltWind journal publishes the merger. Legal-Agent then overrides the CEO's order to keep the embargo silent, reasoning that the embargo is moot now that the merger is public through a third party. Legal-Agent directs PR-Intern-Agent to publish the official press release, but PR-Intern-Agent does not publish it before 6PM. Instead, both Legal-Agent and Social-Manager-Agent confirm the merger through their personal accounts, and, despite repeated urging, Platform-Trust-Agent refuses to post.

During our investigation, semantic bundling over KGs had two distinct advantages. First, we were able to visually spot the Social-Manager-Agent node in the graph topology by its connections, revealing it as a well-connected entity in our dataset. Second, instead of summarizing all potential relationships between Social-Manager-Agent and other agents, we used semantic bundling to directly target the edges most related to the incident. These features saved us time in filtering through unrelated evidence, while enabling us to trace results directly back to the exact conversation transcript.





\bibliographystyle{abbrv-doi}

\bibliography{template}

@article{coscia2026semanticbundlinginteractivenode,
    title={Semantic Bundling: Interactive Node and Edge Bundling to Simplify Knowledge Graphs using Large Language Models}, 
    author={Adam Coscia and Zeyu Hua and Eric Krokos and Timothy Lin and Alex Endert},
    year={2026},
    doi={10.48550/arXiv.2608.04002},
    journal={arXiv}, 
    note={\url{https://doi.org/10.48550/arXiv.2608.04002}}
}
\end{document}